# Improved deep learning based macromolecules structure classification from electron cryo tomograms

Chengqian Che[1], Ruogu Lin[2], Xiangrui Zeng[3], Karim Elmaaroufi[4], John Galeotti[1,*], and Min Xu[3,*]

[1]*Robotics Institute, Carnegie Mellon University, Pittsburgh, 15213, USA.*
[2]*Department of Automation, Tsinghua University, Beijing, 100084, China.*
[3]*Computational Biology Department, Carnegie Mellon University, Pittsburgh, 15213, USA.*
[4]*Electrical and Computer Engineering Department, Carnegie Mellon University, Pittsburgh, 15213, USA.*
[*]*Corresponding authors*

## Abstract

Cellular processes are governed by macromolecular complexes inside the cell. Study of the native structures of macromolecular complexes has been extremely difficult due to lack of data. With recent breakthroughs in Cellular electron cryo tomography (CECT) 3D imaging technology, it is now possible for researchers to gain accesses to fully study and understand the macromolecular structures single cells. However, systematic recovery of macromolecular structures from CECT is very difficult due to high degree of structural complexity and practical imaging limitations. Specifically, we proposed a deep learning based image classification approach for large-scale systematic macromolecular structure separation from CECT data. However, our previous work was only a very initial step towards exploration of the full potential of deep learning based macromolecule separation. In this paper, we focus on improving classification performance by proposing three newly designed individual CNN models: an extended version of (Deep Small Receptive Field) DSRF3D, donated as DSRF3D-v2, a 3D residual block based neural network, named as RB3D and a convolutional 3D(C3D) based model, CB3D. We compare them with our previously developed model (DSRF3D) on 12 datasets with different SNRs and tilt angle ranges. The experiments show that our new models achieved significantly higher classification accuracies. The accuracies are not only higher than 0.9 on normal datasets, but also demonstrate potentials to operate on datasets with high levels of noises and missing wedge effects presented.

## 1 Introduction

As the basic unit of life, cell has always been a fundamental focus of biomedical research. Governed by macromolecules, cellular processes occur over a large length scale. To fully understand the biological processes at different levels, it is essential to gain knowledge of native structures and spatial organizations of macromolecular complexes inside single cells. Due to the lack of data acquisition techniques, little has been known about such knowledge due to lack of suitable data acquisition techniques. Recent breakthroughs in Cellular Electron Cryo Tomography (CECT) imaging technique enables the 3D visualization of macromolecular complex structures and their spatial organizations inside single cells at submolecular resolution and close to their native state [14, 22, 18, 43]. CECT has made possible the discovery of numerous important structural features in prokaryotic cells, eukaryotic cells, and viruses [17, 3, 10, 20]. Therefore, CECT emerges as a very promising tool for systematically studying macromolecular complexes with unprecedented coverage, precision and fatality. In principle, a CECT image contains structural information of all macromolecular complexes inside the field of view. However, the systematic recovery of macromolecular structures from CECT is very difficult due to high degree of structural complexity and practical imaging limitations. Specifically, the densely populated cytoplasm makes a very "crowded" cellular environment for macromolecules. Also, macromolecules are dynamically interacting each other, forming more complex and heterogeneous structures [25]. On the other hand, current technical limitations inherent to the process of structure determination via single-particle cryo-EM require collecting very large data sets – often images of several thousands of macromolecules. It would likely require separating and averaging millions of macromolecules represented by subtomograms, potentially containing hundreds of highly heterogeneous structural classes (a *subtomogram* is a cubic sub-image that contains only one macromolecule). Although advances in data acquisition automation makes it no longer difficult to acquire CECT images containing such amount of macromolecules, existing computational approaches have very limited scalability and discrimination ability, making them incapable of automatic processing such large amount of data.

Given this challenging task, a number of previous works have been done for analyzing macromolecules from CECT data. In [24, 4], template searching based algorithms were proposed to localize macromolecules of known structures from CECT data. In 2013, Briggs et al. reviewed a number of subtomogram averaging methods to resolve structure of macromolecular complexes in situ[7]. In addition, unsupervised classification-based approaches were also developed[e.g. 2, 8, 34, 28, 6]. Even though these methods showed promising macromolecule structure separation and recovery results, the scalability is strictly limited by the intensive computations. Other approaches such as rotation invariant feature[35] and pose normalization[40] were proposed to address the task while reducing the computational complexity. However, these approaches are limited by anisotropic resolution from missing wedge effect, and high level of noise in CECT data.

In order to overcome the limitations mentioned above, recently, we were the ***first*** to propose deep learning based approach [38] for separating particles into structurally homogeneous subgroups through supervised feature extraction using Convolutional Neural Network (CNN). Such approach achieved significantly better separation performance in terms of both accuracy and scalability compared with our previous approaches, showing that deep learning based approach is potentially a very powerful tool for large scale particle separation. However, this proof-of-principle work is only an initial step towards exploring the full power of deep learning based large-scale particle separation. The accuracy of classification needs to be substantially improved for better structural reconstruction performance.

In this paper, we focus on improving deep learning-based separation of particles of macromolecular complexes extracted from CECT images by designing new CNN models. The three CNN models we are proposing include: an extended version of (Deep Small Receptive Field)DSRF3D [38], donated as DSRF3D-v2, a 3D residual block based neural network[19], named as RB3D, and a convolutional 3D(C3D)[32] based model, CB3D. Our experiment shows that new proposed models can achieve significantly better classification performances than our previous best CNN model proposed in [38]. Among them, CB3D has the best performances and yield accuracy close to 0.9 for normal datasets. Our models also show promising classification performance over datasets of extremely low SNR (0.01).

## 2 Method

### 2.1 Convolutional Neural Networks

Serving as a powerful tool, convolutional deep neural networks have been widely used by researchers to resolve challenging tasks in computer vision especially for image classification. Inspired by biological processes, CNN models are composed of stacked layers including an input, an output and multiple hidden layers, which include convolutional, pooling or fully connected layers. By stacking multiple processing layers, more and more image features are learned and extracted as the training proceeds. More specifically, each convolutional layer contains numerous of filters, considered as neurons with different weights. Neurons in this layer are connected to regions of neighboring neurons in the previous layer, donated as receptive field. For instance, a 1D convolution input $x$ with filter size of $2m+1$ will yield an output $y_i = \sum_{j=-m}^{m} w_i x_{i-j}$, where $w_j$ is the $j$th weight of the convolutional filter. An activation function is applied after the convolution layer. Some common activation layers include sigmoid, tanh, the rectified linear unit(ReLU)[15], Leaky ReLU[19] and Maxout[16]. These activation functions take the input and perform certain fixed mathematical operations on it. They are used to accelerate the convergence of the optimization process. For example, ReLU is defined as $o^{ReLU}(x) = max(0, x)$. Next, pooling layers are utilized to reduce the computational costs during the training process by down sampling the data. Two common ways of pooling are calculating the local maximum and average values of the pooling windows. After series of stacked convolutional and pooling layers, a fully connected layer is used to extract more global features. Each unit in fully connected layers, as name suggested, is connected to all units from the previous layer. For instance, given an $i$th input $x_i$, the

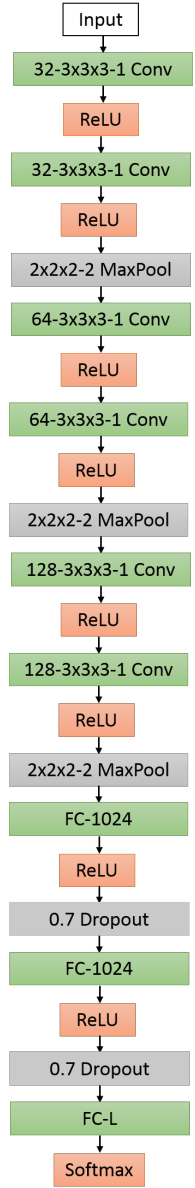

Figure 1: DSRF3D-v2 model: Each box provides configurations for each layer. '32-3x3x3-1 Conv' represents a 3D convolutional layer with 32 5x5x5 filters and stride of 1. 'ReLU' and 'Softmax' are activation layers. '2x2x2-1 MaxPool' means that max operation is implemented over 2x2x2 regions with stride of 2'. 'FC-1024' and 'FC-L' represents fully connected layers with 1024 and L(total number of the classes) neurons respectively.

$j$th output $y_j$ is defined as $y_j = \sum_{i=0}^{n-1} w_{ji} x_i$, where $n$ is the total number of inputs and $w_{ji}$ is the weight between between $i$th input $x$ and $j$th output $y$. Sometimes, special techniques such as Dropout[30] and L2 regularization are used to prevent overfitting. Dropout works by simply only keeping a neuron active with some probability $p$ or setting it to zero otherwise during the training process. Lastly, in order to perform the multi-class classification, a softmax actication function is connected to the last fully connected layer to compute a probability of a sample being assigned to each class. The softmax function is defined in 1, where $f_j(x) = x^T w_j$. $w_j$ are the weights with $j$th class and $P(j|x)$ is the probability of the subtomogram

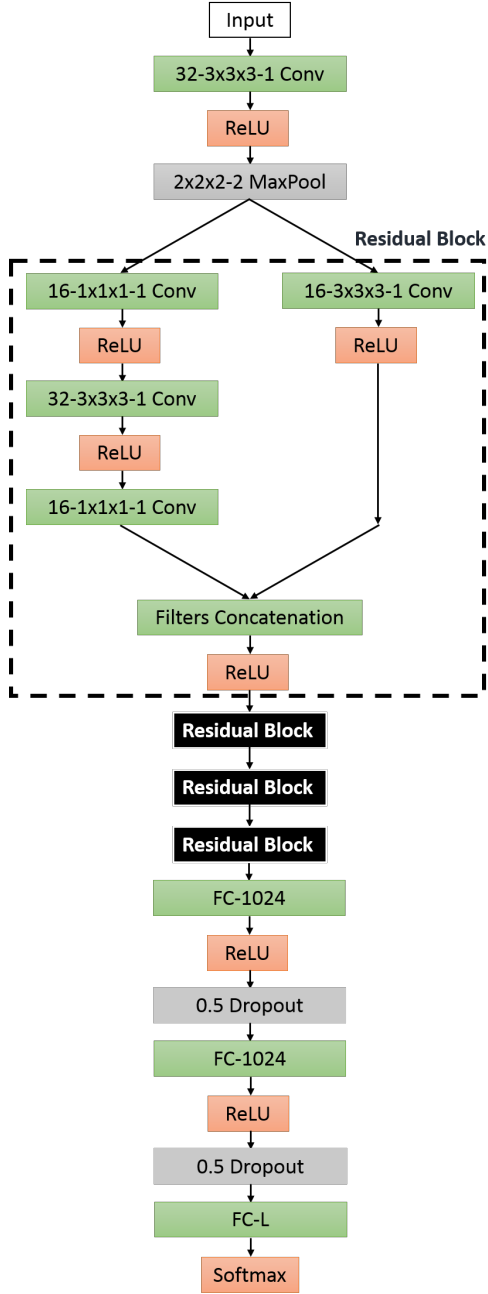

Figure 2: RB3D model: Each box provides configurations for each layer. The definition of the boxes follows Figure 1. Four residual blocks are connected, represented by the black boxes. The specific design for a single residual block is shown in the dashlined box.

is assigned to $j$ class.

$$o_j^{softmax}(x) = P(j|x) = \frac{e^{f_j(x)}}{\sum_{l=1}^{L} e^{f_l(x)}} \qquad (1)$$

Generally speaking, the input data for CNN subtomogram classification is a 3D subtomogram cubic image. The output of the CNN classifier is a vector $o := (o_1, ..., o_L)$, which indicates the probability that a specific subtomogram is predicted to be from each of the $L$ classes.

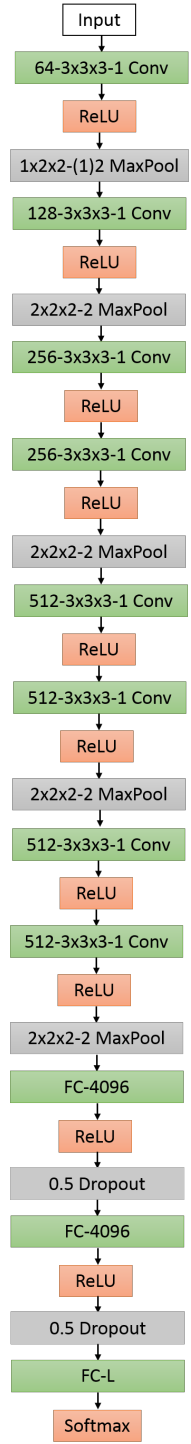

Figure 3: CB3D model: Each box provides configurations for each layer. The definition of the boxes follows Figure 1. Note that '1x2x2-(1)2 MaxPool' means that the max operation is implemented over region of 1x2x2. The stride is 1 for the first dimension and 2 for the others.

Designing CNN architectures and tuning parameters are essential to the performance of networks. In 2012, Krizhevsky et al proposed a novel CNN architecture AlexNet[21], which was the first to show a significant improvement of image classification results on a historically difficult dataset, ImageNet[27]. From that on, CNNs have become a household name in computer vision computer community. In recent years, more advanced CNN architectures were proposed and developed such as GoogleNet[31], ResNet[19] and VGGNet[29]. These networks gradually pushed the classification error rate on ImageNet down to 3.6%[19].

In this paper, we are proposing three different CNN models and comparing them with our previously designed model in [38]. All the models are trained using stochastic gradient descent (SGD) optimizer. We try to minimize the categorical cross-entropy cost function by adding Nesterov momentum of 0.9. In addition, the initial learning rate is set at 0.005 with a decay factor of 1e-7. The training processes are performed with a batch size of 64 for 20 epochs. However, for each dataset, if the classification performance shows no improvement over 5 consective epochs based on the loss function, the training process will end early.

### 2.1.1 DSRF3D-v2 model

In this section, we propose a 3D variant VGGNet[29]-based CNN architecture called Deep Small Receptive Field (DSRF3D). This model is an extended version of our previously proposed model[38] and we donate this model as DSRF3D-v2. Just like a VGGNet, DSRF3D-v2 is featured with sequentially deep stacked layers and small 3D convolution filters with size of 3x3x3. As shown in Figure 1, the input layer is sequentially connected three sets of stacked layers, with each set consisting of 2 3x3x3 3D convolutional layers and one 2x2x2 3D max pooling layer. Then it is followed by two fully connected layers with 70% dropout after each layer. The final fully connected output layer has the same number of units as the structure class number. The activation layers are ReLU for all hidden layers and softmax for fully connected layers. Compared to previously designed DSRF3D model, adding more stacked layers with appropriate dropouts should improve the classification performance intuitively.

### 2.1.2 RB3D model

In this section, a 3-D variant residual block-based[19] CNN model is proposed, donated as RB3D. One big advantage for ResNet-based model is that it avoids negative outcomes while increasing the network overall depth. As shown in Figure 2, this model feeds the input layer to a convolutional layer, a ReLU activation layer and a 2x2x2 3D max-polling layer. Then four bottleneck[19] residual blocks are connected sequentially. For each block, there are two paths that merge together at the end of the block. One path contains one 1x1 layer to reduce dimension, a 3x3 layer and a 1x1 layer for restoring dimension. The other path, considered as a "shortcut", only contains a 3x3 convolutional layer. Lastly, two fully connected layers are constructed with dropout of 50% to prevent overfitting. The usage of residual blocks might lead to higher classification accuracy.

### 2.1.3 CB3D model

In this section, we propose a 3D convolutional (C3D)-based model, named as CB3D. C3D[32] was originally proposed to be trained on large scale supervised video datasets. We can think of the 3D structures of macromolecules as multiple slices of 2D images. If we look from the first slice to the last slice, we can interpret it as a continuously changing object just like a video dataset. In our model shown in Figure 3, we concatenate eight 3D 3x3x3 convolutional layers and each layer is activated by ReLU. Five max pooling layers are mixed among the convolutional layers. At end, two fully connected layers with 50% dropout are added and a softmax activation is appended.

Table 1: The classification accuracy using four CNN models under different SNR and tilt angles

| SNR/Tilt angle range | ±60° | | | | ±50° | | | | ±40° | | | |
|---|---|---|---|---|---|---|---|---|---|---|---|---|
| | DSRF3D | DSRF3D-v2 | RB3D | CB3D | DSRF3D | DSRF3D-v2 | RB3D | CB3D | DSRF3D | DSRF3D-v2 | RB3D | CB3D |
| 0.1 | 0.911 | **0.977** | 0.950 | 0.973 | 0.896 | 0.963 | 0.925 | **0.971** | 0.868 | 0.954 | 0.899 | **0.970** |
| 0.05 | 0.844 | 0.925 | 0.852 | **0.933** | 0.753 | **0.910** | 0.750 | 0.899 | 0.735 | 0.876 | 0.671 | **0.877** |
| 0.03 | 0.706 | 0.841 | 0.711 | **0.849** | 0.581 | 0.746 | 0.548 | **0.747** | 0.537 | 0.038 | 0.502 | **0.717** |
| 0.01 | 0.040 | 0.407 | 0.041 | **0.445** | **0.200** | 0.041 | 0.042 | 0.041 | 0.043 | 0.043 | **0.171** | 0.041 |

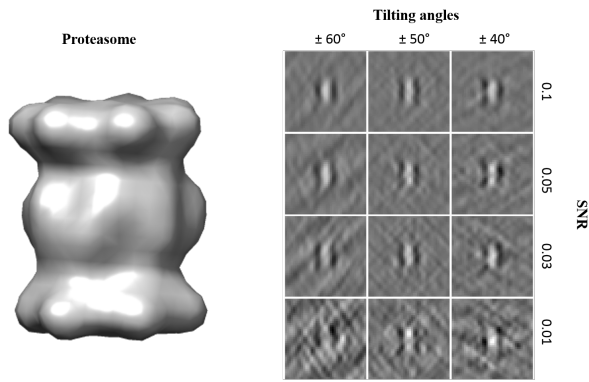

Figure 4: Left: Isosurface of Yeast 20S proteasome (PDB ID: 3DY4); Right: Center slices of subtomograms with different levels of SNRs (0.5, 0.1, 0.05 and 0.01) and tilt angle ranges($\pm 60°$, $\pm 50°$ and $\pm 40°$)

## 2.2 Generation of simulated subtomograms from experimental structures

Similar to previous works [11, 4, 23, 37, 34, 34, 36, 39, 26, 38, 42], we use known structures of macromolecular complexes to generate simulated subtomograms by simulating actual tomographic image reconstruction processes in order to have a reliable assessment of our proposed approaches. There are three significant aspects we focus on when simulating the subtomograms: missing wedge effects, noises, and electron optical factors such as Modulation Transfer Function (MTF) and Contrast Transfer Function (CTF). More specifically, we first generate volumes of $40^3$ voxels with a resolution of 0.92 nm using the PDB2VOL program from the Situs[33] package, and randomly rotate and translate the volumes. Next, we generate projection images of the density maps with different tilt angles to simulate missing wedge effects. The specific tilt angle ranges are $\pm 60°$, $\pm 50°$ and $\pm 40°$. We then convolute the projection images with CTF and MTF [12, 23] to reproduce the electron optical effects to generate simulated electron micrographic images. Then, the simulated noises are added to electron micrographic images [11] with desired signal-to-noise ratio (SNR) levels so that the SNR of reconstructed subtomograms are of 0.1, 0.05, 0.03 and 0.01. The acquisition parameters are set similar to [41], with spherical aberration of 2 mm, defocus of -5$\mu$m, and voltage of 300kV. Finally, with all gathered information, we construct the simulated subtomogram datasets using a direct Fourier inversion reconstruction algorithm implemented in the EMAN2 library [13]. Figure 4 shows an example of center slices of simulated subtomograms with different SNRs and tilt angle ranges.

We construct a simulated dataset for each pair of SNRs(4) and tilt angle ranges(3), which yields 12 sets of data in total. Within a single dataset, for each macromolecular complex, we generate 1000 simulated subtomograms that contain randomly rotated and translated particle of that complex. There are 22 macromolecular complexes collected from the Protein Databank (PDB) [5]. Furthermore, we simulate 1000 subtomograms that contain no macromolecule. As an outcome, each dataset contains 23,000 simulated subtomograms of 23 structural classes.

To fully evaluate classification performances, we first split each dataset into two parts: 80% are used as training data and 20% are used as testing data. Then we take 20% of the training data and use it as validation data for parameter tuning during training process, and the rest 80% for training weights. Therefore, we end up with 14720 training samples, 3680 validation samples and 4600 testing samples. We use the same partitioned datasets across all models to have a fair comparison on their classification performance.

## 2.3 Implementation details

This work is implemented using Keras[9] with Tensorflow[1] as back-end. Keras is a python-based, high-level neural networks API for fast deep learning experimentation. The experiments are performed on a computer with three Nvidia GTX 1080 GPUs, one Intel Core i7-6800K CPU and 128GB memory. The new proposed models are implemented with the same system as our previously proposed model[38].

# 3 Experiment results

## 3.1 Classification performance

In this section, we compare the classification performance of new CNN models(DSRF3D-v2, RB3D and CB3D) with our previously proposed best model DSRF3D[38] on datasets with different SNRs and tilt angles ranges. The results are shown in Table 1. The best performances for each pair of imaging conditions are highlighted in bold. In general, our new proposed CNN models demonstrate significant improvements in classification performance. As shown, the best performance is achieved by newly designed models in all 11 out of 12 situations. The only exception that new models perform worse than DSRF3D is with tilt angle ranges of $\pm 50°$ and SNR of 0.001. Given such poor qualities of images, the "best" performance is only around 0.2, which is still considered as bad performance.

When comparing among the three new models, we can observe that RB3D often end up with lower accuracy than DSRF3D-v2 and CB3D. In fact, RB3D establish obvious improvements (greater than 0.3) only with SNR of 0.1. In situations with low SNRs, RB3D does not show apparent improvement and it performs even worse than DSRF3D. Therefore, the RB3D model is not robust to image noises. In contrast, both DSRF3D-v2 and CB3D show essential accuracy increases in most cases. With SNR of 0.01 and tilt angles of $\pm 60\circ$, both models can remarkably achieve accuracy greater than 0.4. Even though 0.4 is still not considered as high classification accuracy in the Table 1, it at least shows the capability of deep learning to classify a large amount of subtomograms with extremely poor image quality.

DSRF3D-v2 and CB3D generate very similar results, usually with differences less than 0.1. It is proved that both model can accurately classify subtomograms even with much noises and missing wedge effects presented. CB3D is slightly better than DSRF3D-v2 because it achieves the best performance in 8 out of 12 datasets. With SNR of 0.03 and tilt angle ranges of $\pm 40°$, CB3D can still obtain classification accuracy higher than 0.7, which is a very good performance if we consider the poor image qualities.

## 3.2 Classification capability

In this section, we examine the capability of deep learning to classify large scale datasets of subtomograms. We first extract the the best performance for each of the 12 datasets and then plot the highest accuracy with respect to both SNRs and tilt angle ranges. The 3D surface is plotted in Figure 5. As shown, the classification accuracy decreases as more noises are added to the dataset for all tilt angles. Similarly, for all SNRs, the accuracy will reduce if tilt angle ranges decrease from $\pm 60°$ to $\pm 40°$. Based on the plot, we can observe that for datasets whose SNRs are above or equal to 0.05, our best model can achieve classification accuracy no lower than 0.877. For datasets with poor image qualities, as long as SNR is kept above 0.03, classification can achieve higher than 0.7 for all 3 tilt angle ranges. It is proven that our proposed approach has strong abilities to accurately classify macromolecular structures from CECT images and even greater potentials to process datasets with extremely high level of noises and miss wedge effects.

# 4 Conclusions

In this paper, three novel CNN models are proposed to significantly improve deep learning based separation of macromolecules extracted from CECT images. We compare them with our previously proposed model and our best model CB3D ends up with classification accuracy of approximately 0.9 for image datasets with relatively high SNR. More importantly, it demonstrates huge potentials to operate on datasets with extremely poor image qualities. After successfully and efficiently subdividing the subtomograms, the computationally intensive reference-free approaches can be applied to selected subsets separately in order to recover the structure of macromolecular complexes. The overall computational cost can be greatly reduced through such divide and conquer approach. This proof-of-principle work represent a useful step towards full systematic structural separation and recovery of millions of macromolecules extracted from CECT images.

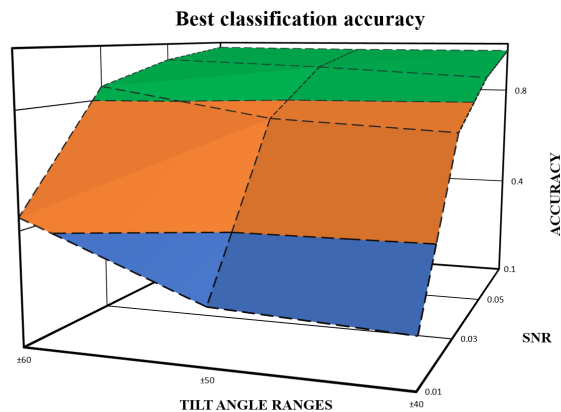

Figure 5: The highest classification accuracy with respect to different SNRs and tilt angle ranges

# 5 acknowledgements

This work was supported in part by U.S. National Institutes of Health (NIH) grant P41 GM103712. J.G. acknowledges support from NIH R01 grant 1R01EY021641, National Library of Medicine contract HHSN27620100058OP and DoD Peer Reviewed Medical Research Program (PR130773, HRPO Log No. A-18237). M.X. and X.Z. acknowledge support of Samuel and Emma Winters Foundation.

# References

[1] Martín Abadi, Paul Barham, Jianmin Chen, Zhifeng Chen, Andy Davis, Jeffrey Dean, Matthieu Devin, Sanjay Ghemawat, Geoffrey Irving, Michael Isard, et al. Tensorflow: A system for large-scale machine learning. *arXiv preprint arXiv:1605.08695*, 2016.

[2] A. Bartesaghi, P. Sprechmann, J. Liu, G. Randall, G. Sapiro, and S. Subramaniam. Classification and 3D averaging with missing wedge correction in biological electron tomography. *Journal of structural biology*, 162(3):436–450, 2008.

[3] M. Beck, V. Lučić, F. Förster, W. Baumeister, and O. Medalia. Snapshots of nuclear pore complexes in action captured by cryo-electron tomography. *Nature*, 449(7162):611–615, 2007.

[4] M. Beck, J.A. Malmström, V. Lange, A. Schmidt, E.W. Deutsch, and R. Aebersold. Visual proteomics of the human pathogen Leptospira interrogans. *Nature methods*, 6(11):817–823, 2009.

[5] H.M. Berman, J. Westbrook, Z. Feng, G. Gilliland, TN Bhat, H. Weissig, I.N. Shindyalov, and P.E. Bourne. The protein data bank. *Nucleic acids research*, 28(1):235, 2000.

[6] Tanmay AM Bharat, Christopher J Russo, Jan Löwe, Lori A Passmore, and Sjors HW Scheres. Advances in single-particle electron cryomicroscopy structure determination applied to sub-tomogram averaging. *Structure*, 23(9):1743–1753, 2015.

[7] John AG Briggs. Structural biology in situ—the potential of subtomogram averaging. *Current opinion in structural biology*, 23(2):261–267, 2013.

[8] Xuanli Chen, Yuxiang Chen, Jan Michael Schuller, Nassir Navab, and Friedrich Forster. Automatic particle picking and multi-class classification in cryo-electron tomograms. In *Biomedical Imaging (ISBI), 2014 IEEE 11th International Symposium on*, pages 838–841. IEEE, 2014.

[9] François Chollet. keras. https://github.com/fchollet/keras, 2015.

[10] Lidia Delgado, Gema Martínez, Carmen López-Iglesias, and Elena Mercadé. Cryo-electron tomography of plunge-frozen whole bacteria and vitreous sections to analyze the recently described bacterial cytoplasmic structure, the stack. *Journal of structural biology*, 189(3):220–229, 2015.

[11] F. Förster, S. Pruggnaller, A. Seybert, and A.S. Frangakis. Classification of cryo-electron sub-tomograms using constrained correlation. *Journal of structural biology*, 161(3):276–286, 2008.

[12] J. Frank. *Three-dimensional electron microscopy of macromolecular assemblies*. Oxford University Press, New York, 2006.

[13] Jesús G Galaz-Montoya, John Flanagan, Michael F Schmid, and Steven J Ludtke. Single particle tomography in eman2. *Journal of structural biology*, 190(3):279–290, 2015.

[14] Lu Gan and Grant J Jensen. Electron tomography of cells. *Quarterly reviews of biophysics*, 45(01):27–56, 2012.

[15] Ian Goodfellow, Yoshua Bengio, and Aaron Courville. *Deep Learning*. MIT Press, 2016. http://www.deeplearningbook.org.

[16] Ian Goodfellow, David Warde-Farley, Mehdi Mirza, Aaron Courville, and Yoshua Bengio. Maxout networks. In Sanjoy Dasgupta and David McAllester, editors, *Proceedings of the 30th International Conference on Machine Learning*, volume 28 of *Proceedings of Machine Learning Research*, pages 1319–1327, Atlanta, Georgia, USA, 17–19 Jun 2013. PMLR.

[17] Kay Grünewald, Prashant Desai, Dennis C Winkler, J Bernard Heymann, David M Belnap, Wolfgang Baumeister, and Alasdair C Steven. Three-dimensional structure of herpes simplex virus from cryo-electron tomography. *Science*, 302(5649):1396–1398, 2003.

[18] Kay Grünewald, Ohad Medalia, Ariane Gross, Alasdair C Steven, and Wolfgang Baumeister. Prospects of electron cryotomography to visualize macromolecular complexes inside cellular compartments: implications of crowding. *Biophysical chemistry*, 100(1):577–591, 2002.

[19] Kaiming He, Xiangyu Zhang, Shaoqing Ren, and Jian Sun. Deep residual learning for image recognition. *arXiv preprint arXiv:1512.03385*, 2015.

[20] Marion Jasnin, Mary Ecke, Wolfgang Baumeister, and Günther Gerisch. Actin organization in cells responding to a perforated surface, revealed by live imaging and cryo-electron tomography. *Structure*, 24(7):1031–1043, 2016.

[21] Alex Krizhevsky, Ilya Sutskever, and Geoffrey E Hinton. Imagenet classification with deep convolutional neural networks. In *Advances in neural information processing systems*, pages 1097–1105, 2012.

[22] Vladan Lučić, Alexander Rigort, and Wolfgang Baumeister. Cryo-electron tomography: the challenge of doing structural biology in situ. *The Journal of cell biology*, 202(3):407–419, 2013.

[23] S. Nickell, F. Förster, A. Linaroudis, W.D. Net, F. Beck, R. Hegerl, W. Baumeister, and J.M. Plitzko. TOM software toolbox: acquisition and analysis for electron tomography. *Journal of Structural Biology*, 149(3):227–234, 2005.

[24] S. Nickell, C. Kofler, A.P. Leis, and W. Baumeister. A visual approach to proteomics. *Nature reviews Molecular cell biology*, 7(3):225–230, 2006.

[25] Long Pei, Min Xu, Zachary Frazier, and Frank Alber. Simulating cryo electron tomograms of crowded cell cytoplasm for assessment of automated particle picking. *BMC Bioinformatics*, 17:405, 2016.

[26] Long Pei, Min Xu, Zachary Frazier, and Frank Alber. Simulating cryo electron tomograms of crowded cell cytoplasm for assessment of automated particle picking. *BMC bioinformatics*, 17(1):405, 2016.

[27] Olga Russakovsky, Jia Deng, Hao Su, Jonathan Krause, Sanjeev Satheesh, Sean Ma, Zhiheng Huang, Andrej Karpathy, Aditya Khosla, Michael Bernstein, et al. Imagenet large scale visual recognition challenge. *International Journal of Computer Vision*, 115(3):211–252, 2015.

[28] S.H.W. Scheres, R. Melero, M. Valle, and J.M. Carazo. Averaging of electron subtomograms and random conical tilt reconstructions through likelihood optimization. *Structure*, 17(12):1563–1572, 2009.

[29] Karen Simonyan and Andrew Zisserman. Very deep convolutional networks for large-scale image recognition. *arXiv preprint arXiv:1409.1556*, 2014.

[30] Nitish Srivastava, Geoffrey E Hinton, Alex Krizhevsky, Ilya Sutskever, and Ruslan Salakhutdinov. Dropout: a simple way to prevent neural networks from overfitting. *Journal of Machine Learning Research*, 15(1):1929–1958, 2014.

[31] Christian Szegedy, Sergey Ioffe, and Vincent Vanhoucke. Inception-v4, inception-resnet and the impact of residual connections on learning. *arXiv preprint arXiv:1602.07261*, 2016.

[32] Du Tran, Lubomir D Bourdev, Rob Fergus, Lorenzo Torresani, and Manohar Paluri. C3d: generic features for video analysis. *CoRR, abs/1412.0767*, 2(7):8, 2014.

[33] W. Wriggers, R.A. Milligan, and J.A. McCammon. Situs: A Package for Docking Crystal Structures into Low-Resolution Maps from Electron Microscopy. *Journal of Structural Biology*, 125(2-3):185–195, 1999.

[34] M. Xu, M. Beck, and F. Alber. High-throughput subtomogram alignment and classification by Fourier space constrained fast volumetric matching. *Journal of Structural Biology*, 178(2):152–164, 2012.

[35] M. Xu, W. Li, G.M. James, M.R. Mehan, and X.J. Zhou. Automated multidimensional phenotypic profiling using large public microarray repositories. *Proceedings of the National Academy of Sciences*, 106(30):12323–12328, 2009.

[36] Min Xu and Frank Alber. Automated target segmentation and real space fast alignment methods for high-throughput classification and averaging of crowded cryo-electron subtomograms. *Bioinformatics*, 29(13):i274–i282, 2013.

[37] Min Xu, Martin Beck, and Frank Alber. Template-free detection of macromolecular complexes in cryo electron tomograms. *Bioinformatics*, 27(13):i69–i76, 2011.

[38] Min Xu, Xiaoqi Chai, Hariank Muthakana, Xiaodan Liang, Ge Yang, Tzviya Zeev-Ben-Mordehai, and Eric Xing. Deep learning based subdivision approach for large scale macromolecules structure recovery from electron cryo tomograms. *ISMB/ECCB 2017, Bioinformatics (in press); arXiv preprint arXiv:1701.08404*, 2017.

[39] Min Xu, Elitza I Tocheva, Yi-Wei Chang, Grant J Jensen, and Frank Alber. De novo visual proteomics in single cells through pattern mining. *arXiv preprint arXiv:1512.09347*, 2015.

[40] Xiao Ping Xu, Christopher Page, and Niels Volkmann. *Efficient Extraction of Macromolecular Complexes from Electron Tomograms Based on Reduced Representation Templates*. Springer International Publishing, 2015.

[41] Tzviya Zeev-Ben-Mordehai, Daven Vasishtan, Anna Hernández Durán, Benjamin Vollmer, Paul White, Arun Prasad Pandurangan, C Alistair Siebert, Maya Topf, and Kay Grünewald. Two distinct trimeric conformations of natively membrane-anchored full-length herpes simplex virus 1 glycoprotein b. *Proceedings of the National Academy of Sciences*, 113(15):4176–4181, 2016.

[42] Xiangrui Zeng, Miguel Ricardo Leung, Tzviya Zeev-Ben-Mordehai, and Min Xu. A convolutional autoencoder approach for mining features in cellular electron cryo-tomograms and weakly supervised coarse segmentation. *arXiv preprint arXiv:1706.04970*, 2017.

[43] Peijun Zhang. Correlative cryo-electron tomography and optical microscopy of cells. *Current opinion in structural biology*, 23(5):763–770, 2013.